\documentclass[cits]{PoS}
\usepackage{slashed}
\usepackage{amsmath}

\title{The role of pions as virtual constituents of light bound states}

\ShortTitle{Pions as virtual constituents of light bound states}

\author{{Walter Heupel}%
         \thanks{Poster Presentation}\\
        Justus-Liebig-Universitaet Giessen\\
        E-mail: \email{walter.heupel@theo.physik.uni-giessen.de}}

\author{Stanislav Kubrak%
         \thanks{Poster Presentation}\\
        Justus-Liebig-Universitaet Giessen\\
        E-mail: \email{stanislav.kubrak@physik.uni-giessen.de.}}

\author{Gernot Eichmann\\
       Karl-Franzens-Universitaet Graz\\
        E-mail: \email{gernot.eichmann@theo.physik.uni-giessen.de}}

\author{\speaker{Christian Fischer}\\
       Justus-Liebig-Universitaet Giessen\\
        E-mail: \email{christian.fischer@theo.physik.uni-giessen.de}}

\abstract{We summarize recent results for the spectrum and properties of light pseudoscalar
and scalar hadrons in a functional approach to QCD using Dyson-Schwinger and Bethe-Salpeter
equations. We focus in particular on the role of pions as virtual constituents in light
mesons ('pion cloud effects') as well as their role as constituents of tetraquark bound
states. An extension of our framework into the charm quark sector reveals an all-charm
tetraquark in the mass region around 5.5 GeV.}

\FullConference{51st International Winter Meeting on Nuclear Physics\\
                 21-25 January 2013\\
                 Bormio (Italy)}

\begin{document}

\section{Introduction}
Pions are the lightest hadrons in nature. Their special status as the lightest members of
an octet of pseudoscalar (pseudo-)Goldstone bosons associated with the breaking of the $SU(3)_A$
part of chiral symmetry makes them fascinating objects to study. In effective theories,
pions appear as elementary degrees of freedom. Naturally, these theories have nothing to
say about the internal structure of the pion. Within QCD, however, pions are formed as
bound states of a quark and an anti-quark interacting strongly with each other. Powerful
constraints such as the axial Ward-Takahashi identity guarantee the massless Goldstone
boson nature of the pions in the chiral limit, although the constituents remain massive.
This fascinating interplay between dynamical mass generation on the quark level and the
formation of massless bound states is particularly transparent in functional approaches
to QCD using Dyson-Schwinger, Bethe-Salpeter and Faddeev equations, see e.g.
\cite{Maris:2003vk,Fischer:2006ub,Maris:2005tt,Eichmann:2008ef} for reviews.

An important probe for the internal structure of hadrons is their response to electromagnetic
fields. Real and virtual photons serve to extract quantities such as form factors,
polarizabilities and distribution functions which in turn offer interesting insights into
global as well as spatially resolved electromagnetic properties of these hadrons. Virtual
pions (the 'pion cloud' \cite{Thomas:1981vc}) play an important role in this respect.
At small momentum transfer, the incoming photons couple dominantly to the virtual pion cloud
of the bound state instead of the quark and antiquark constituents. An impressive
demonstration of this effect has been given in \cite{Eichmann:2011vu}, where the electromagnetic
form factors of the nucleon have been determined in a three-body Faddeev approach. A similar
picture has been found for the axial and pseudoscalar form factors \cite{Eichmann:2011pv}.
The inclusion of pion cloud effects into form factor calculations within the functional approach
is challenging, both conceptually and numerically. First steps in this direction have been made
in \cite{Fischer:2007ze,Fischer:2008wy}, where pion cloud effects in the masses of light mesons
have been studied. In this proceedings contribution we report on our efforts to determine
meson form factors in this approach.

Another interesting place where virtual pions may play a role are tetraquark bound states.
The story of the tetraquark is tightly connected to the issue of the light scalars and
especially to the case of the $\sigma$ meson. Back in the 70ies, Jaffe introduced a simple
quark-bag model to describe these states \cite{Walter:Jaffe1}. He showed that tetraquarks
allow for a natural explanation for the inverted mass spectra and the huge width found in
the lightest $0^+$ meson nonet. Unfortunately this huge width and the multitude of other
particles with $0^+$ quantum numbers  in the low mass region made a clear experimental
signal difficult to obtain. Even the existence as a bound state, described as a pole in
the S-matrix, was debated for decades. Only recently new experimental data coming from
BES \cite{Walter:Bes1} and KLOE \cite{Walter:Cloe1} combined with the application of a modified
Roy-equation \cite{Walter:Leutwyler,Walter:Paelez} technique asserted the existence of the
$\sigma$ meson.

The mass of the {$\sigma$/$f_0(500)$} was found to be $m_\sigma = 450+i280 \mathrm{MeV}$
\cite{Walter:Leutwyler,Walter:Paelez} and the particle has been promoted to a proper member
of the PDG again \cite{Walter:PDG}. While the existence of the $f_0(500)$ seems beyond doubt,
its interpretation in terms of its internal structure is still a matter of debate.
Several approaches, including effective theory and large $N_c$ studies
\cite{Walter:Santopinto,Walter:Achasov,Walter:Black,Walter:Giacosa,Walter:Klempt, Walter:Ebert}
as well as more recent lattice calculations \cite{Walter:Marthur,Walter:Prelovsek}, indicate
a strong contribution from a $qq\bar{q}\bar{q}$-like operator.

In this contribution we report on our study of tetraquarks within an approach using a
covariant four-body equation, which we approximate by a diquark and meson constituent picture.
These are determined self-consistenly from the the underlying quark and gluon substructure employing
a model for the quark-gluon interaction known to reproduce hadron properties
on a phenomenological level~\cite{Walter:Tandy}.


\section{Meson form factors}
\subsection{Dyson-Schwinger and Bethe-Salpeter equations beyond rainbow-ladder}

\begin{figure}
\centering
\includegraphics[scale=0.30]{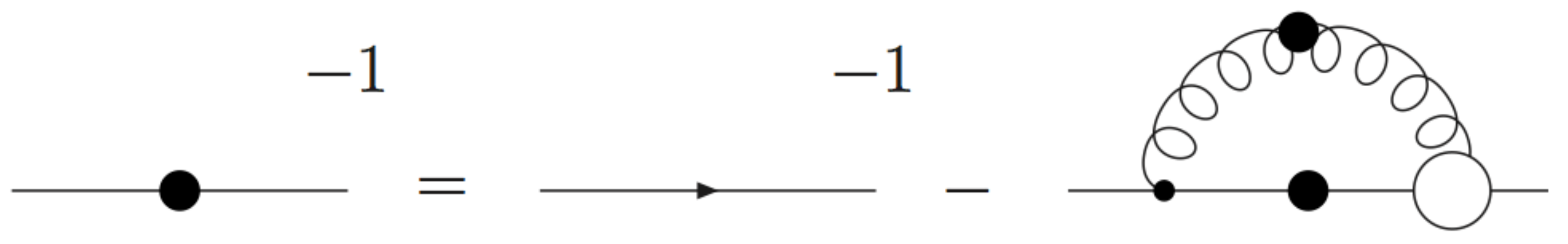}
\caption{The Dyson-Schwinger equation for the fully dressed
quark propagator.}
\label{fig:dse1}
\end{figure}

The Dyson-Schwinger equation (DSE) for the quark propagator is given diagrammatically in
Fig.~\ref{fig:dse1}.
On the left-hand side we find the fully dressed inverse quark propagator
$S^{-1}(p)=i \slashed{p}\, A(p^2) + B(p^2)$. The first term on the right-hand
side denotes its bare counterpart $S^{-1}_{0}(p) = i \slashed{p} + m$, and
the self-energy includes again the fully dressed quark propagator as well
as the gluon propagator and a bare and a dressed quark-gluon vertex.

For phenomenological purposes, a commonly used approximation scheme for the
coupled system of the quark DSE and Bethe-Salpeter
equation (BSE) is the rainbow-ladder approach. In the quark DSE this
approximation amounts to the simple choice of $\Gamma(p,q)_\mu = \gamma_\mu \,\Gamma(k)$
for the quark-gluon vertex, where $p$ and $q$ are the ingoing and outgoing
quark momenta and $k$ is the momentum of the gluon. This approximation ignores
further tensor structures of the full vertex as well as the dependence of the
dressing function on the quark momenta. The remaining dressing $\Gamma(k)$ can
then be combined with the dressing $Z(k)$ of the Landau-gauge gluon
propagator,
\begin{eqnarray}
D_{\mu\nu}(k) = \left(\delta_{\mu \nu}
- \frac{k_\mu k_\nu}{k^2}\right)
\frac{Z(k^2)}{k^2}\,,  \label{gluon}
\end{eqnarray}
into a single effective coupling for the quark-gluon interaction. The most important
property of this approximation is that it readily allows for the construction of a
corresponding interaction kernel in the Bethe-Salpeter equation of mesons, thus
satisfying the axial-vector Ward-Takahashi identity (axWTI). This is mandatory to
obtain the pion in the chiral limit as both, a Goldstone boson and a bound state of
a massive quark-antiquark pair.

\begin{figure}
\includegraphics[scale=0.55]{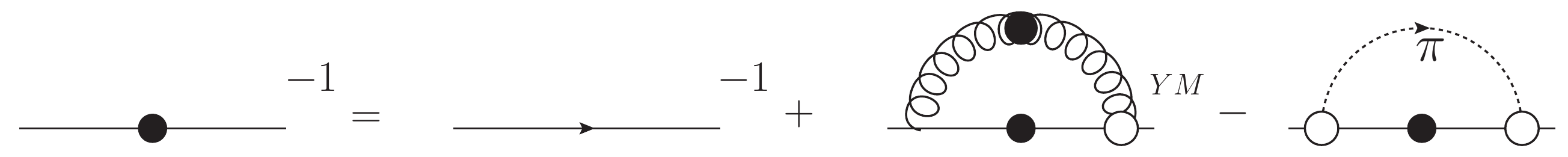}\\
\caption{Dyson-Schwinger equation for quark propagator with explicit diagram for the pion back-reaction.}
\label{DSE2}
\end{figure}

Although there is much phenomenological success in describing mass spectroscopy
of light hadrons, decay constants \cite{RLspectra1,RLspectra2,RLspectra3} and dynamical
properties like meson form factors \cite{RL_formfact} with good agreement to 
experiment~\cite{RL_and_exp}, the rainbow-ladder scheme is not sufficient when it comes
to the inclusion of unquenching effects like pion-cloud contributions. In the tower of
coupled Dyson-Schwinger equations, such contributions appear already in the DSE for
the quark-gluon vertex, see~\cite{Fischer:2008wy} for details. When these contributions
are taken into account explicitly, the resulting quark-gluon interaction can be split into a
pure Yang-Mills part (present already in the quenched approximation) and a separate
part denoting the effects of the back-reaction of pions onto the quark propagator.
The DSE is then given diagrammatically in Fig.~\ref{DSE2} \cite{Fischer:2008wy}.
The corresponding equation is given by
\begin{equation}
\begin{array}{l}
\displaystyle S^{-1}(p)=Z_2 \,S^{-1}_0(p) + g^2 \,C_F (Z_2)^2 \int \frac{d^4 k}{(2\pi)^4} \,\gamma_{\mu} \,S(k) \,\gamma_{\nu} \left(  \delta_{\mu \nu}-\frac{q_\mu q_\nu}{q^2} \right)  \frac{Z(q^2)\,\Gamma_{YM}(q^2)}{q^2} \\[5mm]
\displaystyle - \frac{3}{2} \int \frac{d^4 k}{(2\pi)^4} \left[Z_2 \,\gamma_5 \,S(k) \,\Gamma_\pi \left( \frac{p+k}{2};k-p \right) + Z_2 \,\gamma_5 \,S(k) \,\Gamma_\pi \left( \frac{p+k}{2};p-k \right) \right] \frac{D_\pi(q^2)}{2}\,,
\end{array}
\label{DSE}
\end{equation}
with the momentum routing $q=p-k$ and the pion propagator $D_\pi(q^2)=1/(q^2+M^2_\pi)$.
The product $Z(k^2)\,\Gamma_{YM}(k^2)$ stands for the Yang-Mills part of the quark-gluon interaction,
which we approximate in the spirit of the rainbow-ladder approach. The Maris-Tandy model
\cite{RLspectra3} that parametrizes $Z(k^2)\,\Gamma_{YM}(k^2)$ is given by:
\begin{equation}
\displaystyle Z(k^2)\Gamma_{YM}(k^2)=\frac{4\pi}{g^2} \left( \frac{\pi}{w^6}\
D \,k^4 \,exp(-k^2/w^2) + \frac{2\pi\gamma_m\,[1-exp(-q^2/4m^2_t)]}{log(\tau + (1+k^2/\Delta^2_{QCD}))} \right),
\label{MT_model}
\end{equation}
with $m_t=0.5\:GeV$, $\tau=e^2-1$, $\gamma_m=12/(33-2N_f)$ and $\Delta_{QCD}=0.234\:GeV$.
In the pion part of the interaction, $\Gamma_\pi (p;P)$ is the full pion wave function 
which we approximate by its leading amplitude in the chiral limit \cite{RLspectra1}:
\begin{equation}
\Gamma^j_\pi(p;P)=\tau^j\,\gamma_5\,\frac{B_{\chi}(p^2)}{f_\pi}\,.
\label{Pion_vertex}
\end{equation}
Here, $B_{\chi}(p^2)$ is the scalar dressing function of the quark propagator in the chiral limit.
The replacement of the leading physical pion amplitude by this chiral-limit approximation is
correct on the few-percent level. Note that we use this approximation only for the internal
pion which mediates the interaction.

\begin{figure}[b]
\includegraphics[scale=0.6]{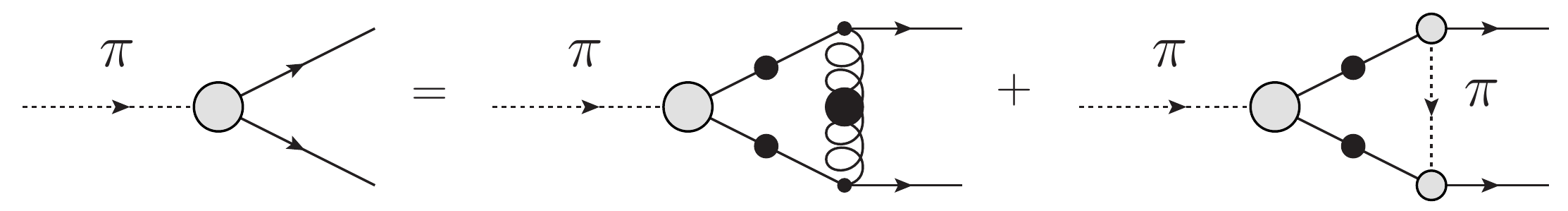}
\caption{Bethe-Salpeter equation for quark anti-quark bound state with explicit diagram for the pion back-reaction.}\label{fig:bse1}
\end{figure}

The Bethe-Salpeter equation for meson bound states is shown diagrammatically in Fig.~\ref{fig:bse1}.
The corresponding equation reads:
\begin{equation}
\displaystyle \Gamma^{(\mu)}_{tu}(p;P)=\int \frac{d^4 k}{(2\pi)^4}
\left( K^{YM}_{tu;rs} (p,k;P) + K^{pion}_{tu;rs}(p,k;P) \right)
\left[ S(k_+)\,\Gamma^{(\mu)}(k;P)\,S(k_-) \right]_{sr}\,,
\label{BSE_1}
\end{equation}
with the kernels $K^{YM}_{tu;rs}$ and $K^{pion}_{tu;rs}$ given by:
\begin{align}
\displaystyle K^{YM}_{tu;rs}(p,k;P)&= \frac{g^2 Z(q^2) \Gamma^{YM}(q^2) (Z_2)^2 }{q^2} \left(  \delta_{\mu \nu}-\frac{q_\mu q_\nu}{k^2} \right) 
\left[ \frac{\lambda^a}{2} \gamma_\mu \right]_{ts} \left[ \frac{\lambda^a}{2} \gamma_\nu \right]_{ru}, \label{BSE_2} \\
\displaystyle K^{pion}_{tu;rs}(p,k;P)&=\frac{1}{4}[\Gamma^j_\pi]_{ru}  \left( \frac{p+k-P}{2};p-k \right)[Z_2 \tau^j \gamma_5]_{ts} \,D_\pi(q^2) \nonumber\\
 &+ \frac{1}{4}[\Gamma^j_\pi]_{ru}  \left( \frac{p+k-P}{2};k-p \right)[Z_2 \,\tau^j \,\gamma_5]_{ts} \,D_\pi(q^2) \label{BSE_3} \\
 &+ \frac{1}{4}[\Gamma^j_\pi]_{ru}  \left( \frac{p+k+P}{2};p-k \right)[Z_2 \,\tau^j \,\gamma_5]_{ts} \,D_\pi(q^2) \nonumber\\
 &+ \frac{1}{4}[\Gamma^j_\pi]_{ru}  \left( \frac{p+k+P}{2};k-p \right)[Z_2 \,\tau^j \,\gamma_5]_{ts} \,D_\pi(q^2) \,. \nonumber
\end{align}
Here, $\Gamma^{(\mu)}(p;P)$ is the Bethe-Salpeter vertex function of a quark-antiquark bound state.
The Latin indexes $(t,u,r,s)$ of the kernels refer to their Dirac structure. It has been shown explicitly in
\cite{Fischer:2007ze} that this interaction kernel satisfies the axWTI.

\subsection{Coupling the meson bound state to an external field}
\begin{figure}
\centering
\includegraphics[scale=0.4]{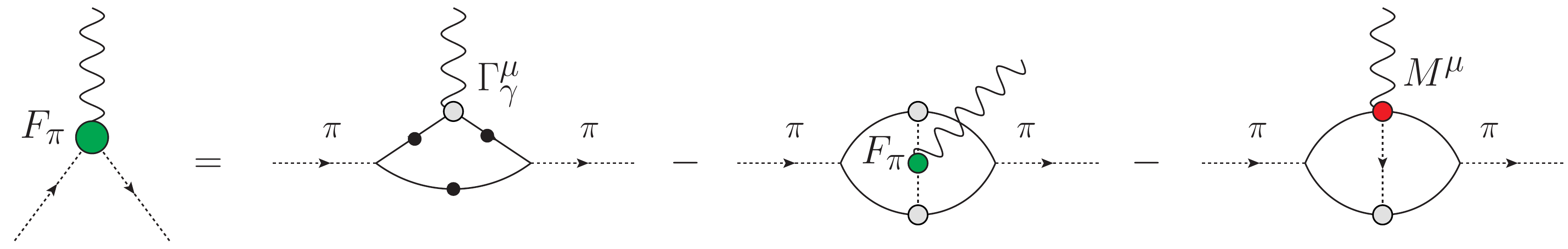}
\caption{The pion form factor. All internal vertexes and propagators are dressed.}
\label{piform}
\end{figure}

A systematic procedure to couple bound states to external gauge fields was given by \cite{Gauging_intro}
and applied to the diquark-quark model of baryons in Ref.~\cite{Oettel:1999gc} and to the
three-body Faddeev equation in Refs.~\cite{Eichmann:2011vu,Eichmann:2012mp}.
In short, the evolution of the two-body quark system is given by the amputated version of the
scattering matrix $T$. This function can be obtained by solving a Dyson equation:
\begin{equation}
\displaystyle T=-iK-iKG_0T\,,
\label{Gauging_1}
\end{equation}
where $G_0$ is the disconnected product of two full quark propagators and $-iK$ is the
two-quark interaction kernel. When the two-quark system forms a bound state, the
scattering matrix develops a pole at $P^2=-M^2$, and can be defined as:
\begin{equation}
\displaystyle T\approx \frac{\Gamma \,\bar{\Gamma}}{P^2+M^2}\,.
\label{Gauging_2}
\end{equation}
Substituting \eqref{Gauging_2} in \eqref{Gauging_1} and keeping only the singular term,
we arrive at the Bethe-Salpeter equation for the two-quark bound state:
\begin{equation}
\displaystyle \Gamma = -iKG_0\Gamma\,.
\label{Gauging_3}
\end{equation}
Then a systematic coupling to the external gauge field gives for $T^{(2)}$:
\begin{equation}
\displaystyle T^\mu = T(iK^{-1} K^\mu K^{-1} + G_0^\mu)T\,.
\label{Gauging_4}
\end{equation}
The bound-state electromagnetic current $J^\mu$ can be expressed at the pole by:
\begin{equation}
\displaystyle T^\mu \approx \frac{\Gamma_f}{P^2_f+M^2_f}J^\mu\frac{\bar \Gamma_i}{P^2_i+M^2_i}\,.
\label{Gauging_5}
\end{equation}
Substituting this in \eqref{Gauging_4} and using \eqref{Gauging_3}, we get:
\begin{equation}
\displaystyle J^\mu= \Gamma_f (-iG_0 K^\mu G_0 + G_0^\mu) \Gamma_i\,.
\label{Gauging_EM_current}
\end{equation}
So far, in rainbow-ladder approximation the first term $K_{RL}^\mu=K_{RL}^{(2),\mu}=0$ 
because the gluon does not couple to a photon. Including the pion back-coupling, however,
the gauged kernel does contribute, since the photon can couple to the exchanged pion.
This fact generates two additional diagrams for the pion form factor, which are given in
Fig.~\ref{piform}.
Here $M^\mu$ is an ansatz for the pion-quark-quark-photon-vertex, which can be built
along the corresponding one for the diquark-quark-quark-photon vertex derived in
\cite{Oettel:1999gc}. It reads:
\begin{equation}
\begin{array}{c}
\displaystyle M^\mu=q_q\frac{(4(p-q)-Q)^\mu}{4(p-q) \cdot Q-Q^2} \left( \Gamma((p-q)-Q/2) - \Gamma(p-q) \right) \\[0.4cm]
\displaystyle + q_{ex}\frac{(4(p-q)-Q)^\mu}{4(p-q) \cdot Q-Q^2} \left( \Gamma((p-q)+Q/2) - \Gamma(p-q) \right)\,.
\end{array}
\label{Seagull_vertex}
\end{equation}
In comparison to rainbow-ladder, the calculation  becomes more complicated. For example,
the second diagram involves the pion form factor itself, so that will lead to the
necessity to perform a self-consistent, iterative calculation  which is also
complicated by the two-loop integration.

\begin{figure}[b]
\centering
\includegraphics[scale=0.5]{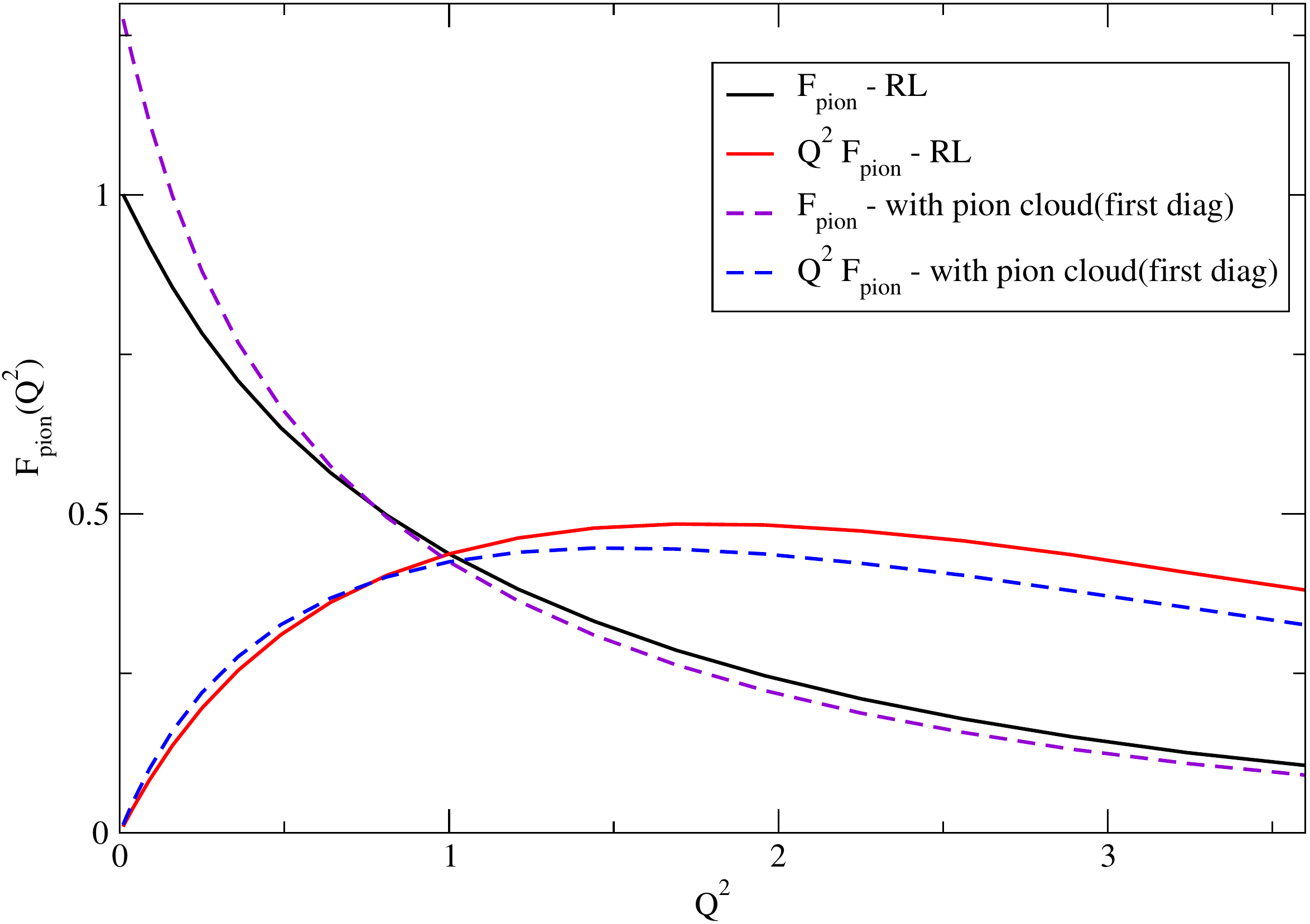}\\
\caption{The comparison of pion form factors within rainbow-ladder truncation and with pion back-coupling included.}
\label{pires}
\end{figure}

Up to now we have performed only a part of the calculation, namely the first diagram
in Fig.~\ref{piform}. The resulting pion form factor with pion back-coupling is shown in Fig.~\ref{pires}.
Although the general behavior of the form factor is already satisfying, we observe sizable deviations
from one at zero momentum transfer. This violation of current conservation stems from the omission of the
pion exchange as well as the seagull diagram in Fig.~\ref{piform} and may indicate the size of corrections
to be expected from these contributions. We are currently in the process of completing the calculation of
the two remaining diagrams. In the future we expect to be able to extend the framework to other light- and
heavy-meson channels as well as into the baryon sector.

\section{Tetraquarks}

\subsection{Bethe-Salpeter equation for the tetraquark}

\begin{figure}[b]
     \centering
      \includegraphics[width=1.0\textwidth]{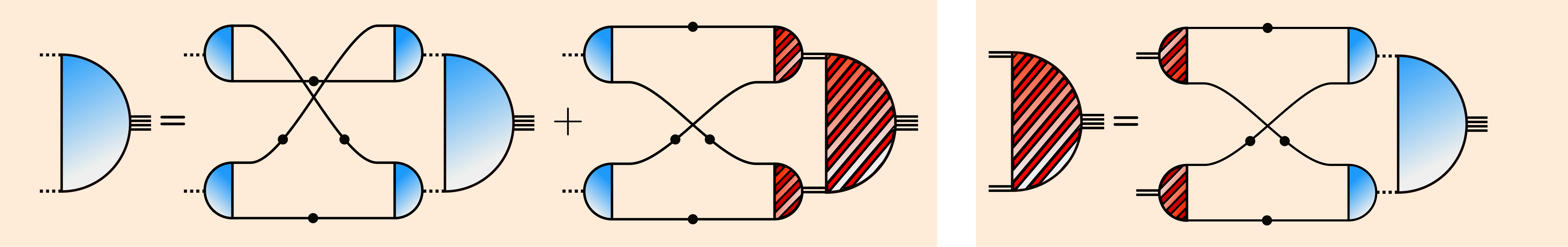}
     \caption{Tetraquark BSE in the meson-meson/antidiquark-diquark picture. The hatched amplitudes involve diquark quantities;
the remaining ones are of mesonic nature. Single (double, dashed) lines are dressed quark (diquark, meson) propagators.}
\label{fig:Tetra_BSE}
\end{figure}
Bound states can be characterized as poles in the appropriate n-point Green function. The amplitude
and the mass of a bound state are found by solving a Bethe-Salpeter equation \cite{Walter:Bethe}.
The general structure of a multi-particle Bethe-Salpeter equation and especially the tetraquark BSE
can be found in \cite{Walter:Huang,Walter:Kvinikidze}. They have the common structure:
\begin{equation}\label{eq:Tetra_BSE}
 \Psi = K G_0\Psi.
\end{equation}
$\Psi$ denotes the amplitude of the tetraquark bound-state, $K$ is the four-body interaction
kernel and $G_0$ represents a product of four fully-dressed quarks. The multiplication is defined
in a functional sense and assumes implicit integration of all intermediate momenta and Dirac,
color and flavor indices. Following the successful strategy in the baryon sector \cite{Walter:Eichmann1},
we drop all 3PI and 4PI contributions. The resulting truncated kernel is a sum of the three
remaining pair interactions \cite{Walter:Kvinikidze}:
\begin{equation}\label{eq:kernel}
 K=\sum_{aa^\prime}K_{aa^\prime}\quad\mathrm{and}\quad K_{aa^\prime}=K_a + K_{a^\prime} - K_aK_{a^\prime}.
\end{equation}
The subscripts $a$ and $a^\prime$ denote $q\bar{q}$, $qq$ and $\bar{q}\bar{q}$ pairs, respectively.
In principle Eq.~(\ref{eq:Tetra_BSE}) can now be solved with the same techniques used for the
nucleon \cite{Walter:Eichmann2}. This is, however, a numerically demanding undertaking and
therefore a reduction to a two-body system is employed. 

The construction of an appropriate 2-body system involves the pair-interacting scattering T-matrix
\begin{equation}\label{eq:T_matrix_pair_interaction}
  T_{aa^\prime}=T_a + T_{a^\prime} + T_aT_{a^\prime}
\end{equation}
 which is closely related to the interaction kernels via Dyson's equation:
\begin{equation}
 T_{aa^\prime}=K_{aa^\prime}(1+T_{aa^\prime})\quad\mathrm{and}\quad T_a=K_a(1+T_a),
\end{equation}
where we drop the first two terms in Eq.~(\ref{eq:T_matrix_pair_interaction}).
It is important to note that the two-body scattering matrix $T_a$ contains a pole of either
a meson or diquark:
\begin{equation}
 T_a=-\Gamma_aD_a\bar{\Gamma}_a,
\end{equation}
where $\Gamma$ is the meson or diquark amplitude and $D$ the corresponding propagator. These
pole contributions are assumed to be dominant.
The key idea in the construction of a two-body equation is now the substitution of the
interaction kernels with the T-matrices in a pole approximation.
This ensures the BSE to be expressed in previously calculated degrees of freedom:
mesons, diquarks and quarks.

Putting everything together, the BSE in Eq.~(\ref{eq:Tetra_BSE})
reduces from a four-body equation, featuring quarks with gluons as exchange particles, to an effective two-body equation.
 The constituents are now mesons and diquarks interacting via quark exchange, see Fig.~\ref{fig:Tetra_BSE}. We take into account only
the mesons and diquarks with lowest mass, expecting the higher-mass state contributions to be subleading. 
It is interesting to note that our approach does not permit a pure diquark-antidiquark state. 
Combining both equations in Fig.~\ref{fig:Tetra_BSE} renders diquark-antidiquark contributions to appear internally only. Thus, one may view the resulting tetraquark bound state as a meson molecule with
diquark-antidiquark admixture to its kernel.

\begin{figure}[b]
     \centering
      \includegraphics[width=1.02\textwidth]{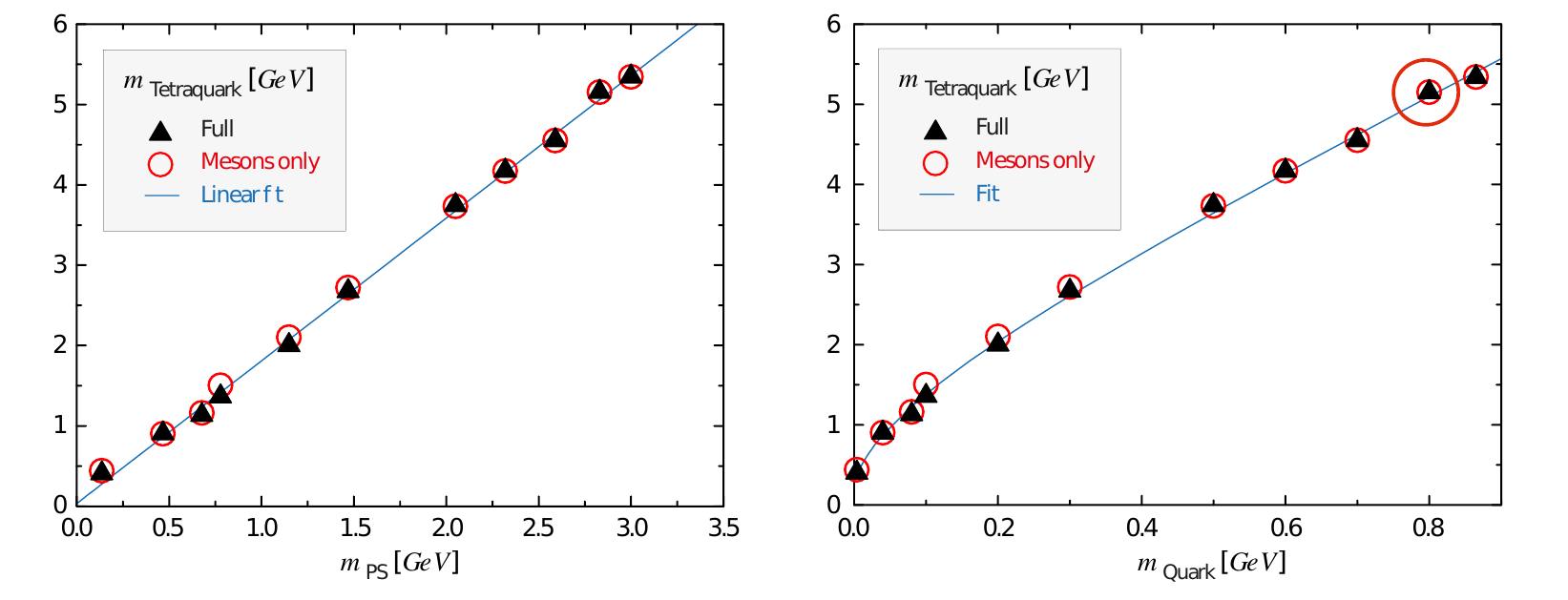}
     \caption{Mass of the $0^{++}$ tetraquark as a function of the pseudoscalar-meson mass (left) and the quark mass \cite{Walter:Walter1}.
              The large red circle highlights a potential all-charm tetraquark.}
     \label{fig:Result}
\end{figure}

\subsection{Results and Discussion}

Our result \cite{Walter:Walter1} for the mass of the up/down $0^{++}$ tetraquark
state as a function of the pseudo\-scalar-meson mass is
shown in the left panel of Fig.~\ref{fig:Result}, together with a calculation that only includes the meson-molecule component
of the tetraquark. The line
represents a fit to the data including a constant, a square
root and a linear term. 
Besides favoring a $\pi\pi$-molecule picture over a subleading diquark-antidiquark component, the main result of our investigation is the value
for the $u/d$ tetraquark at the physical point with $m_{PS}=m_{\pi}$:
\begin{equation}
 m^{u/d}_T(0^{++})=403\, \mathrm{MeV}.
\end{equation}
This value is somewhat lower than the mass deduced from the Roy-equation \cite{Walter:Leutwyler,Walter:Paelez} but other bound states that can mix with the scalar tetraquark are not included in our work. 
The behavior of the tetraquark mass in dependence of the quark mass shows some interesting features: Except for the chiral mass region, the fit resembles the typical behavior of a Goldstone boson. This can be readily understood from
the strongly dominating meson-meson component to the tetraquark bound state. Furthermore, the large decay constant and the low mass of the $\sigma$ is in accordance with our picture of a $\pi\pi$-molecule  
dominated tetraquark.  

It is furthermore interesting to speculate about the
existence of an all-charm tetraquark state.  In Fig.~\ref{fig:Result} the largest pseudoscalar-meson
mass corresponds to a quark mass in the charm region.
We therefore read off the mass of an all-charm scalar
tetraquark state to be at
\begin{equation}
 m^{c}_T(0^{++})=5.3\, \mathrm{GeV}.
\end{equation}
This mass is considerably lower than the 6.2 GeV which were obtained in  simpler model calculations~\cite{Iwasaki:1975pv,Lloyd:2003yc}.
It is also much lower than the $\eta_c$ threshold. Potential decay channels into $D$ mesons and pairs of light
mesons necessarily involve internal gluon lines. This could result in rather small  decay widths. 
Further results for tetraquarks with different quantum numbers are in progress.

\bigskip

\noindent
{\bf Acknowledgements}\\
This work was supported by the Helmholtz International Center for FAIR within the LOEWE program
of the State of Hesse, by the Helmholtz-Zentrum GSI, by BMBF under contract 06GI7121,  
by the Austrian Science Fund FWF under Erwin-Schr\"odinger-Stipendium No. J3039, and by the
DFG transregio TR16.

\end{document}